# Caching Video-on-Demand in Metro and Access Fog Data Centres

Wafaa B. M. Fadlelmula, Sanaa Hamid Mohamed, Taisir E. H. El-Gorashi, and Jaafar M. H. Elmirghani
*School of Electronic and Electrical Engineering, University of Leeds, Leeds, LS2 9JT, UK*
*E-mail: {Elwbf@leeds.ac.uk, elshm@leeds.ac.uk, T.E.H.Elgorashi@leeds.ac.uk, J.M.H.Elmirghani@leeds.ac.uk}*

**ABSTRACT**
This paper examines the utilization of metro fog data centres and access fog datacentres with integrated solar cells and Energy Storage Devices (ESDs) to assist cloud data centres in caching Video-on-Demand content and hence, reduce the networking power consumption. A Mixed Integer Linear Programming (MILP) model is used to optimize the delivery of the content from cloud, metro fog, or access fog datacentres. The results for a range of data centre parameters show that savings by up to 38% in the transport network power consumption can be achieved when VoD is optimally served from fully renewable-powered cloud or metro fog data centres or from access fog data centres with 250 $m^2$ solar cells. Additional 8% savings can be achieved when using ESDs of 100 kWh capacity in the access fog data centres.

## 1. INTRODUCTION

Video-on-demand and live streaming traffic, which are significantly increasing, are expected to account for 82% of the total Internet traffic by 2022 compared to 75% in 2017 [1]. In such services, users expect to get videos streamed smoothly without interruption or delay. To cope with these requirements, some video content providers have migrated to cloud-based services. For instance, Netflix has moved their content fully to Amazon Web Service (AWS), ie Amazon cloud and Open Connect Appliances (OCA) [2]. The latency issue has been managed by using a dedicated Content Delivery Network (CDN), which brings the most popular or requested content closer to the users [3].

The increasing video demands are also accompanied by high data rate traffic especially with the appearance of interactive videos such as 360 degrees video and live stream videos. This can result in increasing the energy consumption of the transport network that carries the traffic to end-users. Reducing this consumption will lead to a reduction in operational costs and ultimately increasing the profit for cloud providers. This can be done by using different techniques for reducing the energy consumption or by greening the data centres by using renewable energy instead of non-renewable energy, utilizing excess heat and by using free natural cooling mechanisms when possible [4]. Several research efforts have focused on developing energy-efficient architectures for cloud data centres and core networks [5] - [8]. To improve the network energy efficiency, different methods have been proposed. These include the optimization of resource allocation where these resources may be virtualized [9] - [13], network architecture and routing design [14] - [21], content distribution optimization [22] - [24], and the optimal processing of big data [25] - [28]. Also the use of renewable energy in core networks was considered in [29]. To alleviate the traffic load on core networks, different distributed architectures have been proposed in [29] - [32]. The authors in [21] minimized the power consumption by developing a Mixed Integer Linear Programming (MILP) optimization model to design IP over WDM networks and found that by applying a lightpath bypass strategy, the power consumption can be reduced by 25% to 45%. In [7], the authors minimized the power consumption by determining the optimal location of a data centre or multiple data centres under IP over WDM with lightpath bypass and non-bypass approaches reaching power saving of up to 37.5%. In addition, several content replication schemes were proposed according to the content popularity, which led to additional power saving of up to 28%. In [24] and [33] the authors optimized the workload and content placement of videos in distributed caches. Furthermore, in [34] the authors evaluated video-on-demand content caching taking into account the cache sizes at each node. They found that 42% of the energy can be saved at different times of the day.

Fog computing plays a complementary role to cloud computing as it fulfils the latency and sensitive requirements of applications by providing part of the cloud services at the edge of the network [35]. It can result in energy saving in the transport network if some applications are served from fog nodes [36]. Connecting micro data centers, with small capacity servers and storage devices to Optical Line Terminal (OLT) in passive optical networks (PON) has been proposed in [37]. The authors in [38] introduced the concept of Nano data centres (NaDa) in the end-users' premises, to share videos, and found that they can save energy by up to 30% and can decrease the network power consumption by bringing the content of some applications closer to users in a peer to peer fashion.

Due to the rapid growth in energy consumption and the increased Greenhouse Gases (GHG), there is a massive need to protect the environment. The transformation into green solutions will minimize the negative effects that influence the environment by reducing the energy consumption and GHG. In [29], the authors sought to reduce the energy consumption by integrating renewable energy into IP over WDM networks which resulted in 47%-52% energy savings with a reduction in $CO_2$ by 97% compared to the peak value and 78% compared to the average value. The authors in [39] showed that deploying virtual machines in nodes that have access to renewable energy resources can result in lowering the carbon emission by 32%. The use of wind farms was considered in [40] while accounting for cloud locations optimization and content replication alongside power losses in transmission.

As a result of the instability of the generated renewable power, as it depends on the unpredictable nature of renewable sources, for example weather, the authors in [41] examined the use of Energy storage devices (ESD) to store energy and hence aid in stabilizing their power supply. In [42] we optimized VoD delivery from cloud or access fog data centres while considering brown and solar energy sources and ESD for the fog data centres. In this paper, we examine the optimization of VoD delivery when data centres placed in the metro network are also considered. The remainder of this paper is organized as follows. Section 2 presents the system model and the parameters used for the MILP model. Section 3 provides the results with discussion while the conclusions and future work are provided in Section 4.

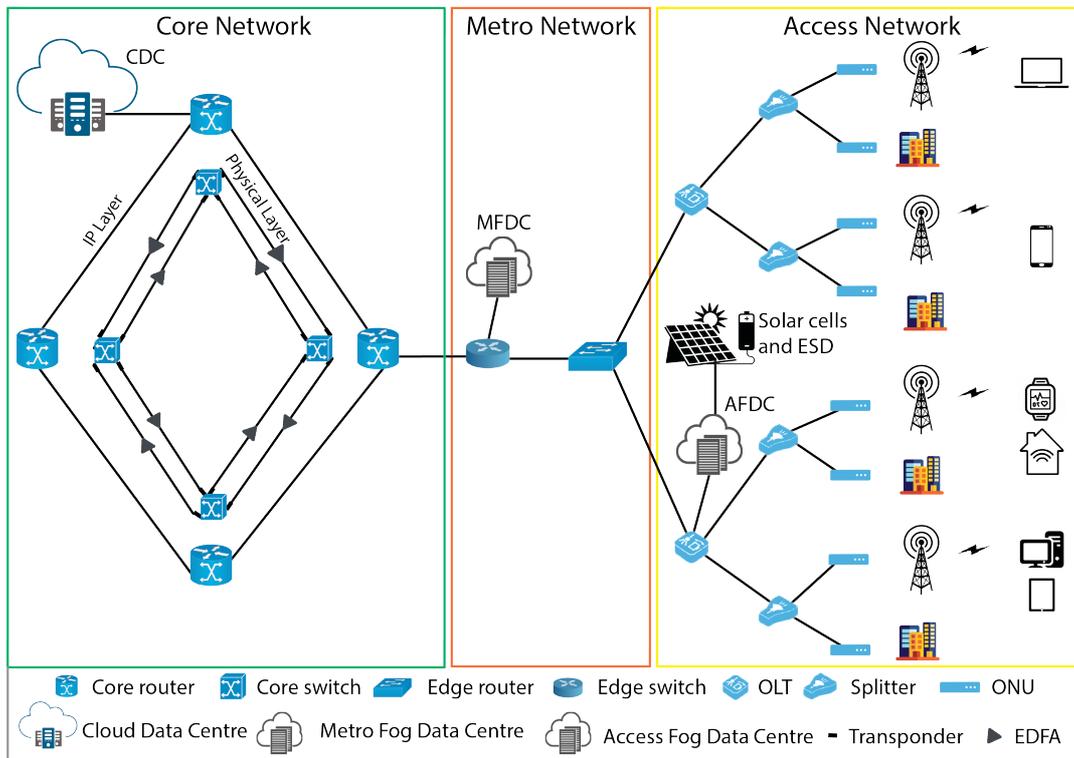

Figure 1: Transport network and cloud-fog data centres architecture

## 2. System model

A MILP model was developed to optimally deliver VoD from cloud data centres (CDC), Metro fog data centres (MFDC) or access fog data centres (AFDC) in an IP over WDM network. The NSFNET topology was utilized for core nodes that house IP routers and optical switches in the physical layer, associated with transponders, and with EDFAs and regenerators placed along fibre routes. We assume that the communication between core nodes follows the lightpath bypass approach [21]. Five CDCs locations were considered and a prediction for VoD demands in 2020 was assumed [5]. Each core node is connected to a metro network which acts as a gateway to access network through edge routers. Passive optical networks (PON) are considered for access networking due to their high bandwidth, reliability and high data transmission rates. Each PON network consists of an OLT, splitters and ONUs to connect end users. Fog data centres are composed of MFDCs which are located in the metro network and AFDCs which are located in the access network. In our setting, each AFDC contains 88 servers with 1.8 Gbps capacity per server. The MFDCs can have up to 5000 servers. Moreover, AFDC are powered by brown energy in addition to renewable energy resources (solar cells) with area of 250 m$^2$ or by stored solar energy in ESDs with capacity of 100 kWh [43]. Table 1. Table 1 summarizes the key parameters for the MILP model including the capacities of links and data centres and the networking equipment power consumption.

*Table 1. Key parameters for the MILP model.*

| Parameter | Value |
|---|---|
| Power consumption of cloud router port (PRc) [12] | 30 Watt |
| Power consumption of metro and access fog router port (PMR & PAR) [12] | 13 Watt |
| Power consumption of cloud and metro fog switch (PCcsw & PCmfsw) [12] | 470 Watt |
| Power consumption of access fog switch (PCafsw) [12] | 210 Watt |
| Power consumption of metro ethernet switch (PCmsw) [12] | 470 Watt |
| Power consumption of an OLT [42] | 904 Watt |
| Cloud, metro fog switch bit rate (Bs) [12] | 600 Gbps |
| Access fog switch bit rate (Bsa) [12] | 240 Gbps |
| Capacity of a content server (Cs) [23] | 1.8 Gbps |
| PUE of cloud data centre ($PUE_c$) | 1.1 |
| PUE of metro fog data centre ($PUE_{MF}$) | 1.2 to 1.1 |
| PUE of access fog data centre ($PUE_{AF}$) | 1.2 to 1.1 |
| Ratio of network equipment to computing power consumption for data centres [42] | 1.3 |
| PUE of core, metro and access networking equipment ($PUE_N$) [5] | 1.5 |
| Total capacity between OLT and access fog data centre [42] | 160 Gbps |
| Total capacity between OLT and metro network [42] | 160 Gbps |
| Total number of metro data centres servers | 5000 |
| Size of a solar cell [42] | 250 m$^2$ |
| Battery maximum capacity ($E_{max}$) [43] | 100 kWh |
| Charging percentage per hour and discharging percentage per hour [41] | 72.25% and 90.25% |

## 3. RESULTS

*A: PUE Impact with Brown-powered Data Centres:*
To evaluate the impact of the optimum VoD delivery, different power usage effectiveness (PUE) values for brown-powered CDC, MFDC and AFDC, were considered in the proposed model. Figure 2 illustrates the total brown power consumption (PC) per day for delivering VoD demands with different combinations of PUE for metro and access fog data centres (i.e $PUE_{MF}$ and $PUE_{AF}$, respectively) while assuming that cloud data centres PUE (i.e. $PUE_{C)}$ is 1.1. The results show that with PUE 1.1 for all data centres, it is more efficient to stream from AFDC. As $PUE_{AF}$ increases and $PUE_{MF}$ is 1.1 it starts to stream from MFDCs until it reaches its full capacity then it streams from CDCs for the remaining demands.

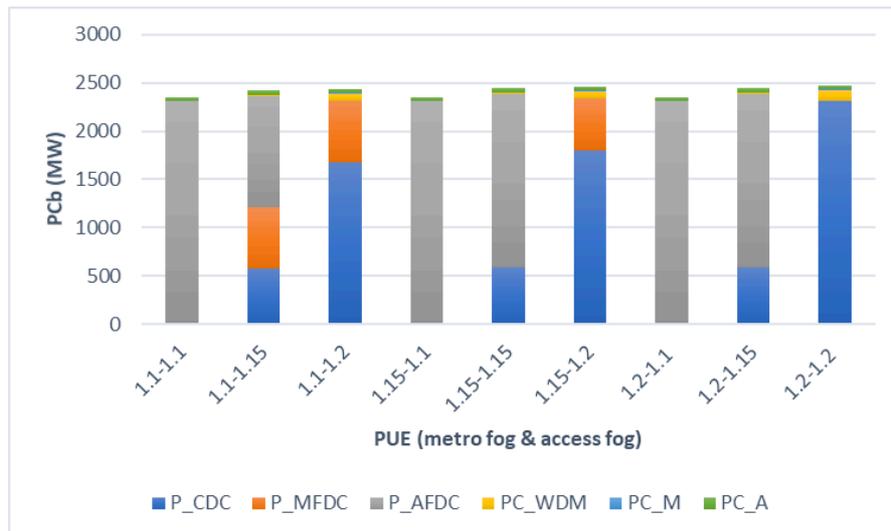

*Figure 2: power consumption under different $PUE_{MF}$ and $PUE_{AF}$ values*

*B. power consumption with fully renewable-powered CDCs and MFDCs and solar- powered AFDC:*
In this case, fully renewable-powered CDCs and MFDCs are considered with $PUF_C$ of 1.1 and $PUE_{MF}$ of 1.1. AFDCs have $PUE_{AF}$ of 1.1 with integrated solar cells size of 250 m$^2$ to power the AFDCs. The results indicate that power savings of up to 38% in the transport network can be achieved compared to the case when fully streaming from brown-powered CDCs. This happens because VoD is served from AFDC during the day hours when solar energy is used and partially served from MFDC which reduces the need to serve from CDCs and consequently reduces the transport network power consumption.

*C. power consumption with fully renewable-powered CDCs and MFDCs with solar-powered AFDC with ESDs:*
Considering ESDs with 100 kWh capacity, in addition to solar cell size of 250 m$^2$ in the case of PUE$_{AF}$ 1.1 can achieve additional savings of up to 8% in the transport network relative to the case without ESDs which is more limited by the sun availability. In this instance, solar power can be saved in batteries which increases the saving. Figure 3 below shows the networking power consumption when streaming from brown-powered CDCs, compared to streaming from renewable-powered CDCs and MFDCs, and AFDC with solar cell size of 250 m$^2$ and finally when ESDs are also considered in the AFDC.

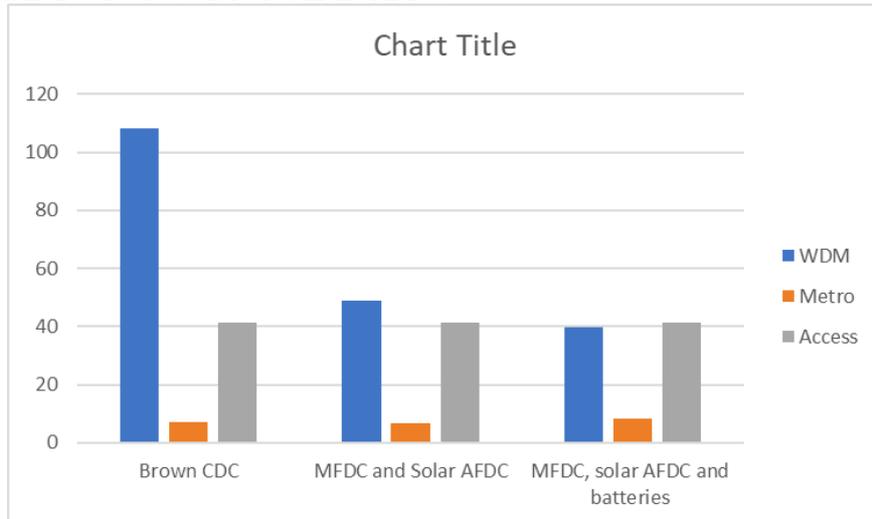

*Figure 3: Network power consumption with different data centres configuration*

### 4. CONCLUSIONS

In this paper, a MILP model was utilized to optimize the delivery of VoD from cloud, metro fog and access fog data centres with solar cells and ESDs with the objective of reducing the brown power consumption. In the first scenario considered, the results show that under brown powered data centres and when all PUEs values are equal to 1.1, it is more efficient to deliver from AFDC. When PUE$_{AF}$ increases and PUE$_{MF}$ is higher than the PUE$_c$ it is more efficient to partially serve from MFDCs and move to cloud after that. In the second scenario, both CDCs and MFDCs are assumed to be fully renewable-powered and solar cells with size of 250 m$^2$ are integrated into the AFDCs. Savings by up to 38% in the transport network can be achieved. Finally, an additional 8% power saving can be achieved by considering ESDs with a capacity of 100 kWh in the access fog data centres. Future work includes considering the impact of popularity of content on the VoD delivery from the different types of data centres.

### ACKNOWLEDGEMENTS
The authors would like to acknowledge funding from the Engineering and Physical Sciences Research Council (EPSRC) INTERNET (EP/H040536/1), STAR (EP/K016873/1) and TOWS (EP/S016570/1) projects. All data are provided in full in the results section of this paper.

### REFERENCES
1. *Cisco Predicts More IP Traffic in the Next Five Years Than in the History of the Internet* 2018.
2. V. Veillon, C. Denninnart and M. A. Salehi, "F-FDN: Federation of Fog Computing Systems for Low Latency Video Streaming," *2019 IEEE 3rd International Conference on Fog and Edge Computing (ICFEC)*, Larnaca, Cyprus, 2019, pp. 1-9.
3. H. Moustafa, E. M. Schooler and J. McCarthy, "Reverse CDN in Fog Computing: The lifecycle of video data in connected and autonomous vehicles," *2017 IEEE Fog World Congress (FWC)*, Santa Clara, CA, 2017, pp. 1-5.
4. C. Cai, L. Wang, S. U. Khan and J. Tao, "Energy-Aware High Performance Computing: A Taxonomy Study," *2011 IEEE 17th International Conference on Parallel and Distributed Systems*, Tainan, 2011, pp. 953-958.

5. J. M. H. Elmirghani *et al.*, "GreenTouch GreenMeter Core Network Energy-Efficiency Improvement Measures and Optimization," *J. Opt. Commun. Netw.*, vol. 10, no. 2, p. A250, 2018.
6. M. O. I Musa, T. E. H El-Gorashi, J. M. H Elmirghani, and S. Member, "Bounds on GreenTouch GreenMeter Network Energy Efficiency," *J. Light. Technol.*, vol. 36, no. 23, p. 5395-5405, 2018.
7. X. Dong, T. El-Gorashi, and J. M. H. Elmirghani, "Green IP over WDM networks with data centers," *J. Light. Technol.*, vol. 29, no. 12, pp. 1861–1880, 2011.
8. H. M. Mohammad Ali, T. E. H. El-Gorashi, A. Q. Lawey, and J. M. H. Elmirghani, "Future Energy Efficient Data Centers With Disaggregated Servers," *J. Light. Technol.*, vol. 35, no. 24, pp. 5361–5380, 2017.
9. L. Nonde, T. E. H. El-Gorashi, and J. M. H. Elmirghani, "Energy Efficient Virtual Network Embedding for Cloud Networks," *J. Light. Technol.*, vol. 33, no. 9, pp. 1828–1849, 2015.
10. A.N. Al-Quzweeni, A. Lawey, T.E.H. El-Gorashi, and J.M.H. Elmirghani, "Optimized Energy Aware 5G Network Function Virtualization," *IEEE Access*, vol. 7, pp. 44939 - 44958, 2019.
11. Z. T. Al-Azez, A. Q. Lawey, T. E. H. El-Gorashi, and J. M. H. Elmirghani, "Energy Efficient IoT Virtualization Framework With Peer to Peer Networking and Processing," *IEEE Access*, vol. 7, pp. 50697–50709, 2019.
12. H. A. Alharbi, T. E. H. El-Gorashi and J. M. H. Elmirghani, "Energy Efficient Virtual Machine Services Placement in Cloud-Fog Architecture," *2019 21st International Conference on Transparent Optical Networks (ICTON)*, Angers, France, 2019, pp. 1-6.
13. H. A. Alharbi, T. E. H. Elgorashi, A. Q. Lawey and J. M. H. Elmirghani, "The Impact of Inter-Virtual Machine Traffic on Energy Efficient Virtual Machines Placement," *2019 IEEE Sustainability through ICT Summit (StICT)*, Montréal, QC, Canada, 2019, pp. 1-7.
14. X. Dong, T. E. H. El-Gorashi, and J. M. H. Elmirghani, "On the energy efficiency of physical topology design for IP over WDM networks," *J. Light. Technol.*, vol. 30, no. 12, pp. 1931–1942, 2012.
15. B. G. Bathula, M. Alresheedi, and J. M. H. Elmirghani, "Energy Efficient Architectures for Optical Networks," *Communications*, Proc IEEE London Communications Symposium, London, pp. 5–8, 2009.
16. B. G. Bathula and J. M. H. Elmirghani, "Energy efficient Optical Burst Switched (OBS) networks," *IEEE GLOBECOM'09, Honolulu, Hawaii, USA, November 30-December 04,*, 2009.
17. T. E. H. El-Gorashi, X. Dong, and J. M. H. Elmirghani, "Green optical orthogonal frequency-division multiplexing networks," *IET Optoelectron.*, vol. 8, no. 3, pp. 137–148, 2014.
18. X. Dong, A.Q. Lawey, T.E.H. El-Gorashi, and J.M.H. Elmirghani, "Energy Efficient Core Networks," *Proc 16th IEEE Conference on Optical Network Design and Modelling (ONDM'12)*, 17-20 April, 2012, UK.
19. M. Musa, T. El-Gorashi, and J. Elmirghani, "Bounds for Energy-Efficient Survivable IP Over WDM Networks With Network Coding," *J. Opt. Commun. Netw.*, vol. 10, no. 5, p. 471-481, 2018.
20. M. Musa, T. El-Gorashi, and J. Elmirghani, "Energy efficient survivable IP-Over-WDM networks with network coding," *J. Opt. Commun. Netw.*, vol. 9, no. 3, pp. 207–217, 2017.
21. G. Shen and R. Tucker, "Energy-minimized design for IP over WDM networks," *IEEE/OSA Journal of Optical Communications and Networking*, vol. 1, no. 1, pp. 176-186, Jun, 2009.
22. A. Lawey, T. E. H. El-Gorashi, and J. M. H. Elmirghani, "BitTorrent content distribution in optical networks," *IEEE/OSA Journal of Lightwave Technology*, vol. 32, no. 21, pp. 3607-3623, 2014.
23. A. Q. Lawey, T. E. H. El-Gorashi, and J. M. H. Elmirghani, "Distributed energy efficient clouds over core networks," *J. Light. Technol.*, vol. 32, no. 7, pp. 1261–1281, 2014.
24. N. I. Osman, T. El-Gorashi, L. Krug, and J. M. H. Elmirghani, "Energy efficient future high-definition TV," *Journal of Lightwave Technology*, vol. 32, no. 13, pp. 2364-2381, Jul. 2014.
25. A. M. Al-Salim, A. Lawey, T. E. H. El-Gorashi, and J. M. H. Elmirghani, "Energy efficient big data networks: Impact of volume and variety," *IEEE Transactions on Network and Service Management*, vol. 15, no. 1, pp. 458-474, 2018.
26. M. S. Hadi, A. Lawey, T. E. H. El-Gorashi, and J. M. H. Elmirghani, "Patient-centric cellular networks optimization using big data analytics," IEEE Access, vol. 7, pp. 49279 - 49296, 2019.
27. A. M. Al-Salim, T. E. H. El-Gorashi, A. Q. Lawey and J. M. H. Elmirghani, "Greening big data networks: velocity impact," in *IET Optoelectronics*, vol. 12, no. 3, pp. 126-135, 6 2018.
28. M. S. Hadi, A. Lawey, T. E. H. El-Gorashi, and J. M. H. Elmirghani, "Big data analytics for wireless and wired network design: A survey," *Elsevier Computer Networks*, vol. 132, no. 2, pp. 180-199, 2018.
29. X. Dong, T. El-Gorashi, and J. Elmirghani, "IP over WDM networks employing renewable energy sources," *Journal of Lightwave Technology*, vol. 29, no. 1, pp. 3-14, Jan. 2011.
30. A. A. Alahmadi, A. Q. Lawey, T. E. H. El-gorashi, and J. M. H. Elmirghani, "Distributed Processing in Vehicular Cloud Networks," in *8th International Conference on the Network of the Future (NOF)*, 2017, pp. 22–26.
31. A. A. Alahmadi, M. O. I. Musa, T. E. H. El-gorashi, and J. M. H. Elmirghani, "Energy Efficient Resource Allocation in Vehicular Cloud Based Architecture," in *2019 21st International Conference on Transparent Optical Networks (ICTON)*, 2019, pp. 1–6.
32. S. H. Mohamed, T. E. H. El-Gorashi and J. M. H. Elmirghani, "Energy Efficiency of Server-Centric PON Data Center Architecture for Fog Computing," *2018 20th International Conference on Transparent Optical Networks (ICTON)*, Bucharest, 2018, pp. 1-4.

33. N. I. Osman, T. El-Gorashi, and J. M. H. Elmirghani, "The impact of content popularity distribution on energy efficient caching," in *Proc. 15th International Conference on Transparent Optical Networks (ICTON)*, Jun. 2013.
34. N. I. Osman, T. El-Gorashi and J. M. H. Elmirghani, "Reduction of energy consumption of Video on-Demand services using cache size optimization," *2011 Eighth International Conference on Wireless and Optical Communications Networks*, Paris, 2011, pp. 1-5.
35. F. Bonomi, R. Milito, J. Zhu and S. Addepalli, "Fog computing and its role in the internet of things". in *Proceedings of the first edition of the MCC workshop on Mobile cloud computing* 2012.
36. F. Jalali, K. Hinton, R. Ayre, T. Alpcan, and R. S. Tucker, "Fog computing may help to save energy in cloud computing," *IEEE Journal on Selected Areas in Communications*, vol. 34, no. 5, pp. 1728-1739, May 2016.
37. B. Yang, Z. Zhang, K. Zhang, and W. Hu, "Integration of micro data center with optical line terminal in passive optical network," in *Proc. 21st OptoElectronics and Communications Conference (OECC)* held jointly with *International Conference on Photonics in Switching (PS)*, Jul. 2016.
38. V. Valancius, N. Laoutaris, L. Massoulie, C. Diot, and P. Rodriguez, "Greening the Internet with nano data centers," in *Proc. 5th International Conference on Emerging Networking Experiments and Technologies, (CoNEXT)*, New York, USA, 2009, pp. 37-48.
39. L. Nonde, T. E. H. El-Gorashi, and J. M. H. Elmirghani, "Virtual network embedding employing renewable energy sources," in *Proc. IEEE Global Communications Conference (GLOBECOM)*, Dec. 2016.
40. A. Q. Lawey, T. E. H. El-Gorashi, and J. M. H. Elmirghani, "Renewable energy in distributed energy efficient content delivery clouds," in *Proc. IEEE International Conference on Communications (ICC)*, Jun. 2015, pp. 128-134.
41. C. Gu, K. Hu, Z. Li, Q. Yuan, H. Huang, and X. Jia, "Lowering down the cost for green cloud data centers by using ESDs and energy trading," in *Proc. IEEE Trustcom/BigDataSE/ISPA*, Aug. 2016, pp. 1508-1515.
42. M. B. Abdull Halim, S. Hamid Mohamed, T. E. H. El-Gorashi and J. M. H. Elmirghani, "Fog Assisted Caching Employing Solar Renewable Energy for Delivering Video on Demand Service," *2019 21st International Conference on Transparent Optical Networks (ICTON)*, Angers, France, 2019, pp. 1-5.
43. H. Chen, T. N. Cong, W. Yang, C. Tan, Y. Li, and Y. Ding, "Progress in electrical energy storage system: A critical review," *Progress in Natural Science*, vol. 19, no. 3, pp. 291-312, 2009.